\documentclass[envcountsame,oribibl]{llncs}
\usepackage{subfigure}
\usepackage{ulem}
\usepackage{url}
\usepackage{amssymb}
\usepackage{amsfonts}
\usepackage{amsmath}
\usepackage{dsfont} 
\usepackage{lipsum}
\usepackage{graphicx}

\usepackage{multirow}
\usepackage{threeparttable}
\usepackage{stfloats}
\usepackage{floatrow}
\usepackage{algorithm}
\usepackage{algorithmic}
\usepackage{adjustbox}
\usepackage{bm}
\usepackage{booktabs}
\usepackage{threeparttable}
\usepackage{caption}
\usepackage{bbding}
\usepackage[misc]{ifsym}
\usepackage{cite}

\usepackage{subfigure}
\usepackage[colorlinks,
            linkcolor=blue,
            anchorcolor=blue,
            citecolor=blue]{hyperref}

\urldef{\mailsa}\path|{alfred.hofmann, ursula.barth, ingrid.haas, frank.holzwarth,|
\urldef{\mailsb}\path|anna.kramer, leonie.kunz, christine.reiss, nicole.sator,|
\urldef{\mailsc}\path|erika.siebert-cole, peter.strasser, lncs}@springer.com|

\begin{document}
\floatsetup[table]{capposition=top}
\mainmatter  

\title{Direct Quantification for Coronary Artery Stenosis Using Multiview Learning}


\author{Dong Zhang$^{1,2}$, Guang Yang$^{3,4}$, Shu Zhao$^{1,2}$, Yanping Zhang$^{1,2}$,\\ Heye Zhang$^5$, Shuo Li$^{6}$}

\institute{ $^1$ Key Laboratory of Intelligent Computing and Signal Processing, Ministry of Education, Anhui University, Hefei, China \\ 
$^2$ School of Computer Science and Technology, Anhui University, Hefei, China \\
$^3$ Cardiovascular Research Centre, Royal Brompton Hospital, London SW3 6NP, UK \\
$^4$ National Heart \& Lung Institute, Imperial College London, London SW7 2AZ, UK\\
$^5$ School of Biomedical Engineering, Sun Yat-Sen University, Shenzhen, China\\
$^6$ Western University, London, ON, Canada
}
\authorrunning{}

\tocauthor{}
\maketitle


\begin{abstract}
The quantification of the coronary artery stenosis is of significant clinical importance in coronary artery diseases diagnosis and intervention treatment. It aims to quantify the morphological indices of the coronary artery lesions such as minimum lumen diameter, reference vessel diameter, lesion length and these indices are the reference of the interventional stent placement. In this study, we propose a direct multiview quantitative coronary angiography (DMQCA) model as an automatic clinical tool to quantify the coronary artery stenosis from X-ray coronary angiography images. The proposed DMQCA model consists of a multiview module with two attention mechanisms, a key-frame module and a regression module, to achieve direct accurate multiple-index estimation. The multi-view module comprehensively learns the spatio-temporal features of coronary arteries through a three-dimensional convolution. The attention mechanisms of each view focus on the subtle feature of the lesion region and capture the important context information. The key-frame module learns the subtle features of the stenosis through successive dilated residual blocks. The regression module finally generates the indice estimation from multiple features. We evaluate the proposed model over 2100 X-ray coronary angiography images collected from 105 subjects from two viewpoints. Compared to other direct quantification methods, our DMQCA model achieves more accurate quantification, enabling to provide a patient-speciﬁc assessment of coronary artery stenosis.
\end{abstract}

\section{Introduction}

The quantification of the coronary artery stenosis (e.g., multiple-index estimation of diameters and lengths for vessel) is of significant clinical importance \cite{zhang2019deep} in coronary artery diseases (CAD) diagnosis and intervention treatment. In particular, the minimum lumen diameter (MLD), reference vessel diameter (RVD) and lesion length (LL) are the most valuable indices for the quantification of the coronary artery stenosis. The MLD and the RVD correlate with the severity of stenosis and the diameter stenosis have an important impact on the blood flow. The LL is crucial for the selection of appropriate stent size in coronary intervention. The accurate index estimation can increase the assessment capabilities for both diagnostic and interventional cardiology.
However, current visual estimation or manual measurement from cardiologists exist differences and errors due to the subjective judgments \cite{Wan}. 
Thus, the direct accurate quantification of the coronary artery stenosis from the X-ray coronary angiography is therefore highly in demand. It is more efficient, accurate, reliable and reproducible compared with visual estimation and manual measurement of the stenosis indices. Existing studies have only focused on the artery lesion grading \cite{Wan}, which only bring a rough description rather than an accurate quantification of the stenosis. Moreover, other studies have performed the quantitative coronary angiography (QCA) with a complex reconstruction \cite{YangJ} first, which can provide more vascular geometry information of the coronary arterial tree but is still not a direct quantification.

\begin{figure}[htbp]
    \centering
    \includegraphics[scale=0.27]{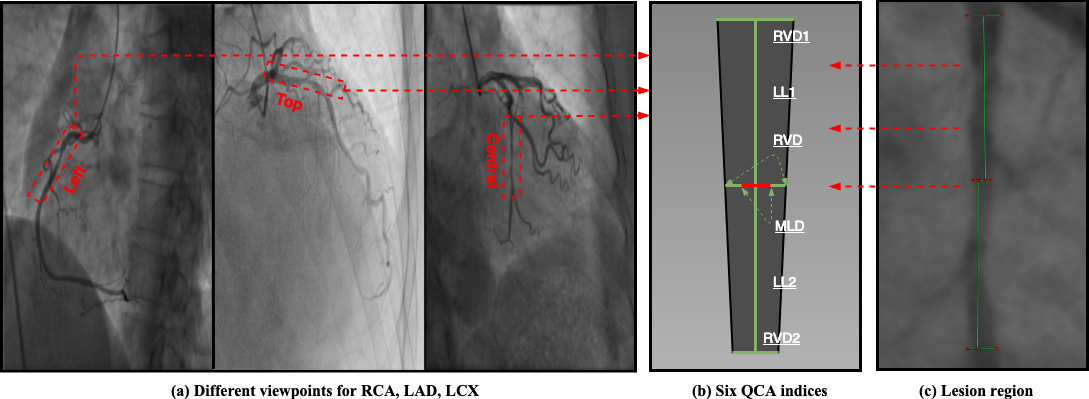}
    \caption{The complex structures and variable locations of coronary arteries in X-ray coronary angiography images, and the lesion regions are relatively small. (a) A $46^{\circ}$ left anterior oblique (LAO) viewpoint for the right coronary artery (RCA). A $44^{\circ}$ right anterior oblique (RAO), $26^{\circ}$ cranial (CRA) viewpoint for the left anterior descending artery (LAD). A $32^{\circ}$ right anterior oblique, $20^{\circ}$ caudal viewpoint (CAU) for the left circumflex coronary artery (LCX). The lesion regions are annotated in red-dashed box, the red coloured texts denote the locations and the red-dashed arrows denote modeling the lesion region as an ideal isosceles trapezoid. (b) The coronary artery is modeled as an ideal isosceles trapezoid and 6 quantitative indices are shown, including four vessel diameters (RVD1, RVD, MLD, RVD2) and two lesion lengths (LL1, LL2). Note that ${RVD}=(RVD2\times LL1+RVD1\times LL2)/(LL1+LL2)$. (c) A zoomed-in region of the coronary artery segment stenosis.}
    \label{fig1}
\end{figure}

Direct quantification of the coronary artery stenosis poses great challenges in feature representation learning. First, the complex structures and various locations of vessels in X-ray images (as shown in Fig. \ref{fig1} (a)). Coronary arteries have many extremely small branches and different shapes due to the individual differences. The severe occlusions always exist and coronary arteries lie in different locations of X-ray images due to the different viewpoints. Because of these, it is difficult to capture the expressive feature representation of different coronary arteries. Second, the artery stenosis is relatively small in the whole X-ray image (as shown in Fig. \ref{fig1} (a)). It is quite challenging to have a comprehensive observation when the angiogram images have high noise, poor contrast and non-uniform illumination. However, these lead to difficulties in capturing the discriminative features of the small lesion objects without segmentation. 
In this paper, a direct multiview \cite{chen2018multiview,yang2018multiview} quantitative coronary angiography model (DMQCA) is proposed to quantify the artery stenosis in X-ray coronary angiography images. This quantification model includes a multiview (main-view, support-view) module with attention mechanisms, a key-frame module and a regression module. Our DMQCA model imitates viewing procedure of the reporting clinicians who make a comprehensive observation based on a main viewpoint, if necessary, a support viewpoint is also employed. As show in Fig. \ref{fig2}, the main-view module and the support-view module are comprised of successive 3D convolution (Conv) networks and integrate self-attention module and context attention module. In particular, the 3DConvs can extract the expressive spatio-temporal features of coronary artery over different time steps from 2D+T sequential X-ray images in each viewpoint. The self-attention \cite{Zhang} module models relationships between widely separated spatial regions in the feature maps. The context attention \cite{HAN} can extract such regions that are important to the current image and such images that are important to the current view, respectively. Then it aggregates the representation of informative regions and images to form the ultimate representation of the image and view. The key-frame module consists of several dilated residual blocks, which are used to capture the subtle features for the stenosis. From the multiview modules and the key-frame module, the expressive spatio-temporal features and the subtle features are aggregated into the regression module, i.e., fully connected network, which is used to capture the relationship between the features and the multiple quantification indices.

The major contribution of this study lies in that we for the first time achieve a direct quantification of coronary artery stenosis from the X-ray coronary angiography images via deep learning. In DMQCA model, (1) we design a multiview learning model, which imitates viewing routines of the reporting clinicians to provide an expressive feature representation for quantitative measurement of the coronary artery stenosis, and (2) introduce two attention mechanisms, which are employed for the model to focus on the important features, especially for extracting the subtle features for artery stenosis.

\section{Methods}

\begin{figure}[htbp]
    \centering
    \includegraphics[scale=0.25]{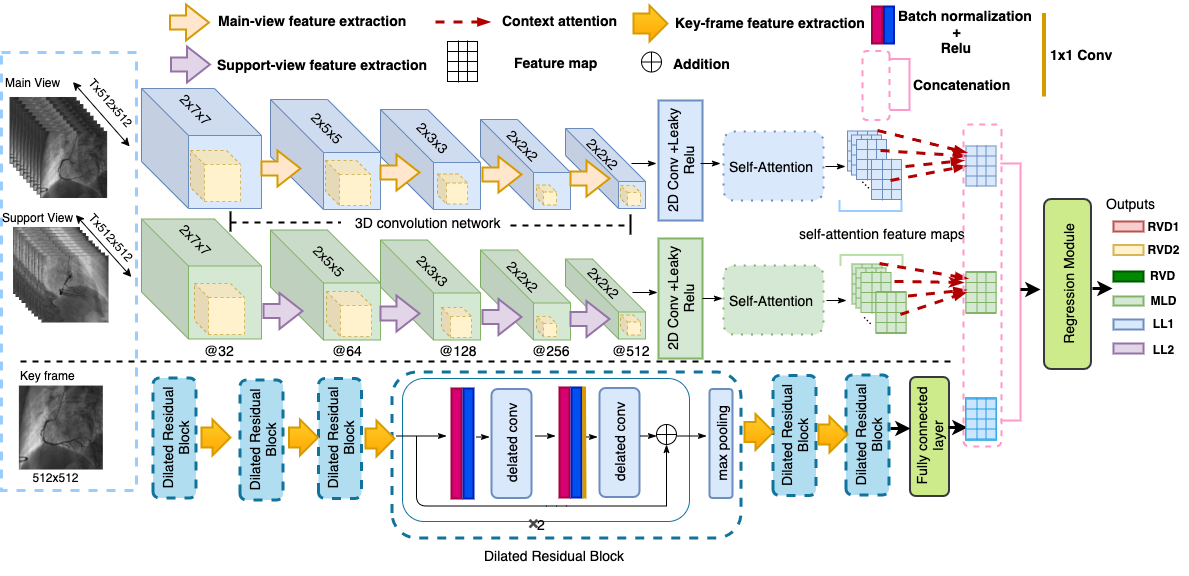}
    \caption{The workflow of the DMQCA model. 3D convolution networks in each view are employed for extract multiview features of coronary artery. $@\#$ denotes the number of 3D convolution filters. The dilated residual block is employed for the feature in key frame. A self-attention and a context attention are employed for the subtle features of the stenosis and the important context information. A regression module is employed for the relationship between the quantitative indices and the expressive features.}
    \label{fig2}
\end{figure}

The proposed DMQCA model is comprised of a multiview module with two attention mechanisms, a key-frame module and a regression module, to achieve direct multiple-index estimation. The workflow of our DMQCA model is summarized as Fig. \ref{fig2}. 

\subsection{{\bf DMQCA Formulation}}

Direct quantification for the coronary artery stenosis is described as a multi-output regression problem solved by our DMQCA model. Consider multiple quantitative indices  $Y$=\{$y_1$,$y_2$,$y_3$,...,$y_d$\} and the objective of the DMQCA is to estimate $Y$ from X-ray coronary angiography images $X$, which consist of two 2D+T image frames $X_m, X_s \in (x_1,x_2,...,x_T, \mathbb{R}^{H\times W \times T}$) of the main viewpoint and the support viewpoint, and a keyframe image $x_{key} \in \mathbb{R}^{H\times W}$, where H and W are the height and width of each frames respectively (H=W=512), T is the temporal step (T=10). Given a training dataset $D=\{X_{im},X_{is},x_{ikey},Y_i\}_{i=1}^{N}$, we aim to learn the mapping $f$ : $X \rightarrow Y \in \mathbb{R}^{d}$, and $N$ is the number of training samples, $d$ is the number of quantitative indices.

\subsection{{\bf Comprehensive Observation for Quantification }}
The DMQCA model can mimic the reporting clinicians to make a comprehensive observation based on a main viewpoint, a support viewpoint and a key frame image. It can learn an expressive feature embedding of coronary arteries stenosis directly from X-ray coronary angiography images.

\subsubsection{{\bf Multiview Feature Embedding Learning.}}
A multiview learning module we design to learn an expressive feature for coronary arteries stenosis. It consists a main-view module and a support-view module, which use the same network architecture as shown in Fig. \ref{fig2}. 3DConvs networks are effective for learning the spatial and temporal features from 3D data. Interestingly, 2D+T image sequences can also be considered as 3D data. In our work, five successive 3DConv layers are designed for each view to extract the morphology (spatial) and kinematic (temporal) features from different temporal steps for coronary arteries. The input of 3DConvs is a set of T image frames. We set the different numbers and the different sizes of 3D convolutional filters (as shown in Fig. \ref{fig2}) with the same $1\times2\times2$ stride.

In order to capture the subtle features of the stenosis in each view, we apply the outputs of the Conv networks into a self-attention block, which models the relationships between the lesion region and the remote regions. In this block, the single image features $x\in R^{C\times M}$ of the frame sequences in current view from the previous layer are first transformed into two feature spaces $f, g$ with different weights, where $M$ is the number of spatial locations, $C$ is the channel numbers, $f(x)=W_{f}x, g(x)=W_{g}x$. In addition, $\alpha _{j,i}$ indicates the contribution weight from the $i_{th}$ location region when synthesizing the $j_{th}$ location. The final output of the self-attention block is denoted as $o=(o_1,o_2,...,o_i,...,o_M)\in R^{C\times M}$, where,
\begin{equation}
    S_{ij}=f(x_i)^{\top} g(x_i)
\end{equation}

\begin{equation}
    \alpha_{j,i}=\frac{exp(S_{ij})}{\sum_{i=1}^{M}exp(S_{ij})}
\end{equation}

\begin{equation}
    o_i=x_i+\gamma\sum_{i=1}^{M}\alpha _{j,i}h(x_i)
\end{equation}
In the formulations above, $W_f, W_g\in R^{\frac{C}{8}\times C}$, $W_h\in R^{C\times C}$ are the weight matrices which are implemented as $1\times1$ convolutions, and the parameter $\gamma$ is initialized as 0.

However, not all regions contribute equally to the representation of each frame meaning. Hence, we introduce the context attention to extract such regions that are important to the meaning of the frame image. Then we aggregate the representation of those regions to form an expressive feature vector of the frame. That is, we first feed the $r_{th}$ region annotation $x_{tr}$ through a one layer perception to get $u_{tr}$ as the hidden feature of $x_{tr}$, then we measure the importance of the region with a region context vector $u_r$ and get a importance weight $\beta_{tr}$ through a softmax function. After that, we compute the $t_{th}$ frame representation $x_t$ as a weighted sum of the regions based on $\beta_{tr}$. Similarly, we again use the context attention mechanism to reward the frames that are important to the current view representation. A frame level context vector $u_f$ is employed to measure the importance $\beta_t$ of the $t_{th}$ frame. Finally, the vector $v$ denotes the view representation that summarizes the information of $T$ frames in the current view. (Here we omit the equations of $u_t, \beta_t, v$ for simplicity.)

\begin{equation}
    u_{tr}=tanh(W_{tr}x_{tr}+b_{tr})
\end{equation}

\begin{equation}
    \beta_{tr}=\frac{exp(u_{tr}^{\top}u_r)}{\sum_{r}^{}exp(u_{tr}^Tu_r)}
\end{equation}

\begin{equation}
    x_t=\sum_{r}^{}\beta_{tr}x_{tr}
\end{equation}

\subsubsection{{\bf Keyframe Feature Embedding Learning.}}
A keyframe module we introduce to enhance the representation of the relatively small stenosis in different locations. The keyframe module consists of 6 successive dilated residual blocks, as shown in Fig. \ref{fig2}. Each dilated residual block employs two residual blocks \cite{He} having two convolution layers with $3\times3$ filters, and a pooling layer. Then we use a fully connected layer to map the feature from dilated residual blocks to the same shape of the multiview modules.

\subsubsection{{\bf Regression Module for the QCA Indices.}}
A regression module we design to aggregate multiple features for direct index estimation. We denote $V_m$, $V_s$, $K$ as the extracted features from the main view, the support view and the keyframe image, respectively. In order to regress the QCA indices according to the multiple features, $V_m, V_s, K$ are concatenated and then we feed it into two fully connected layers with 512 and 6 units, each employing a LeakyReLU activation. The output of the regression module is $f(x_i)=W_o(V_m \oplus V_s \oplus K)+b_o$, where the $\oplus$ denotes the concatenation operator, $W_o$ and $b_o$ are the weight matrix and bias respectively. To minimize the difference between the outputs $f(x_i)$ and the ground truth, we employ the mean absolute error (MAE) as the loss function (Eq. \ref{eq7}), where the $\lambda_{qca}$ is the $l_2$ norm regularization.
\begin{equation}
    L_{qca}=\frac{1}{d\times N}\sum_{i=1}^{N}\left |f(x_i)-Y_i  \right | + \lambda_{qca}\sum_{i}^{}{\left \|w_i  \right \|_{2}}^{2}
    \label{eq7}
\end{equation}

\section{Experiments and Results}
\subsection{{\bf Dataset and Configurations}}
A total of 105 patients of type A coronary artery lesions were retrospectively selected and they have completed coronary angiography using a clinical angiographic X-ray system (Philips Allura XPER). In particular, different severity of lesions locates main coronary artery branches, including 33 LADs, 26 LCXs and 46 RCAs. T=10 frames X-ray images ($512 \times 512$ pixels) are collected from two viewpoints for each patient. The keyframe images are extracted from main viewpoint, and the ground truth values are manually measured under the guidance of an experienced radiologist.

Our deep learning model was implemented using $TensorFlow\ 1.11.0$ on a Ubuntu 16.04 machine, and was trained and tested on an NVIDIA Titan Xp 12GB GPU. For the implementation, we used the Adam method to perform the optimization with decayed learning rate (the initial learning rate was 0.0002) and the $l_2$ norm regularization $\lambda_{qca}$ was set to $10^{-6}$. In our experiments, a 10-fold cross-validation is employed to provide an unbiased estimation of the MAE. It is of note that all the 2100 images are resized into $256\times256$. For comparison studies, the Pearson correlation coefficient is used to evaluate the performance with two baseline methods CNNs, 3DCNNs and other direct quantification methods HOG+RF \cite{Zhen} (histogram of oriented gradients (HOG) feature we use), Indices-Net \cite{Xue} and DMTRL \cite{Xue2}.

\begin{table}[t]

\scriptsize

\caption{Our DMQCA model works best in comparison with two baseline methods, three direct quantification methods and different sub-frameworks including single view (Main, Sup) integrating keyframe (Key) information, main view, keyframe, main view without context attention (Main-ConAtt) and DMQCA without self-attention (\textbf{Ours}-SelfAtt).}
\begin{threeparttable}
\centering
\begin{tabular}{ccccccccc}
\toprule
                  Method              & MAE   & Pearson(\%)   &RVD1   &RVD2   & RVD    & MLD    & LL1    & LL2    \\ \hline
Sup+Key           & $1.3216 \pm 0.7036$ & $88.62 \pm 12.73$ &$0.8642$ &$0.6902$   & $0.7252$ & $0.6642$ & $2.4468$ & $2.5389$ \\ 
Main+Key              & $1.3296 \pm 0.6906$ & $88.40 \pm 11.93$ &0.7780    &0.6735   & 0.8137 & 0.6275 & 2.4768 & 2.6079 \\ 
Key                        & $1.3188 \pm 0.6724$ & $88.72 \pm 12.22$ &0.7864    &0.7082   & 0.6895 & \bm{$0.5648$} & 2.5605 & 2.6032 \\ 
Main-ConAtt &$1.3449 \pm 0.7480$  &$87.61 \pm 14.87$&0.8713        &0.7538        &0.7393        &0.5936        &2.4111        &2.7000        \\ 
Main &$1.3352 \pm 0.6916$   &$87.89 \pm 14.41$   &0.7998        &0.6967        &0.7162        &0.6418        &2.3867        &2.7705        \\ 
\textbf{Ours}-SelfAtt    & $1.3138 \pm 0.6910$ & $87.71 \pm 12.77$   & 0.7913 & 0.7409 & 0.6879 & 0.6258 & 2.5013 & 2.5353 \\ 
$\dagger$ CNNs &$1.3436 \pm 0.6417$   &$87.97 \pm 13.75$   &0.7661        &0.8067        &0.7057        &0.6510        &2.5664        &2.5659        \\

$\$$ 3DCNNs &$1.3537 \pm 0.7035$        &$88.18 \pm 11.53$        &0.7489        &0.7342        &0.7506        &0.6014    &2.5985  &2.6887        \\
$\ddagger$ 3DCNNs &$1.3291 \pm 0.6778$        &$87.86 \pm 12.55$        &0.8143       &0.6694        &0.7881        &0.6744 &2.3800 &2.6482        \\
$\dagger$ HOG+RF \cite{Zhen} &$1.4606 \pm 0.6798$        &$87.47 \pm 14.04$        &0.7915        & 0.7114       &0.7186        &0.5943&2.9396 &3.0086        \\ 
$\dagger$ Indices-Net \cite{Xue} &$1.5105 \pm 0.7475$        &$85.45 \pm 14.40$        &0.8402        &0.7951        &0.9090        &0.7459  &2.7612   &3.0114        \\ 
$\ddagger$ DMTRL \cite{Xue2} &$1.3117 \pm 0.7403$        &$88.44 \pm 12.03$        &\bm{$0.7353$}       &\bm{$0.6591$}        &0.6984        &0.5669  &2.4350   &2.7755        \\ 
$\$$ DMTRL \cite{Xue2} &$1.3124 \pm 0.7362$        &$88.32 \pm 14.31$        &0.7356        &0.6823        &0.6841        &0.6140  &2.4160   &2.7422    \\\hline
\textbf{Ours}                           & \bm{$1.2737 \pm 0.7115$} & \bm{$89.14 \pm 11.24$}   & 0.7947 & 0.7486 & \bm{$0.6669$} & 0.5669 & \bm{$2.3330$} & \bm{$2.5322$} \\ \toprule
\end{tabular}

\begin{tablenotes}
        \scriptsize
        \item[$\dagger$] On Keyframe images.   $\ddagger$  On Main view images. $\$$ On Support view images.  
      \end{tablenotes}
    \end{threeparttable}
\label{t1}
\end{table}

\begin{figure}[htbp]
\setlength{\abovecaptionskip}{-14ex} 
\centering
\subfigure[]{
\begin{minipage}[t]{0.33\linewidth}
\centering
\includegraphics[width=1.6in]{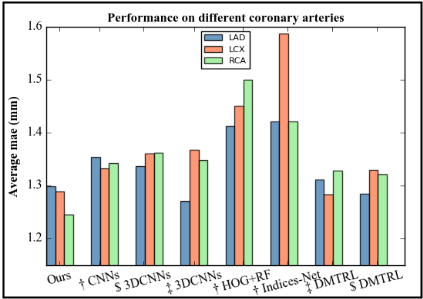}
\end{minipage}%
}%
\subfigure[]{
\begin{minipage}[t]{0.67\linewidth}
\centering
\includegraphics[width=2.9in]{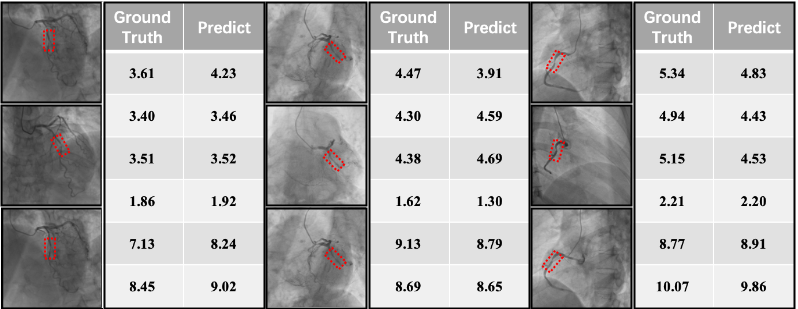}
\end{minipage}%
}%
\centering
\caption{(a) The experimental results of comparison methods on different coronary arteries. (b) The lesion regions are annotated in red-dashed box in main-view, support-view and keyframe (from top to bottom) images. The six indices RVD1, RVD2, RVD, MLD, LL1, LL2 (from top to bottom) of LAD, LCX, RCA (from left to right) quantified by our method are consistent with the ground truth.}
\label{fig3}
\end{figure}

\subsection{{\bf Results and Analysis}}

\subsubsection{Quantitative Coronary Angiography.}
The experimental results show that our DMQCA model can accurately quantify the coronary artery stenosis. The obtained MAE and the Pearson correlation coefficient are $1.2737 \pm 0.7115$ mm and $89.14\% \pm 11.24\%$, respectively. For each quantitative indice, our method estimates the RVD, MLD, LL1 and LL2 with the superior average MAE of 0.6669 mm, 0.5669 mm, 2.3330 mm and 2.5322 mm. 

\subsubsection{Ablation Study \uppercase\expandafter{\romannumeral1}: Multiview Learning.}
We further demonstrate the value of our multiview model for capturing expressive representation. We design some sub-frameworks which just employ single view (main view or support view) with Keyframe (Sup+Key, Main+Key), single view (Main), Keyframe (Key), respectively. The quantitative results in Table \ref{t1} show that our multiview learning achieved better performance for quantitative coronary angiography in X-ray images. 

\subsubsection{Ablation Study \uppercase\expandafter{\romannumeral2}: Attention Mechanism.}We employ a self-attention to force more attention to the lesion regions and apply a context attention to capture the important context information to enhance the representation of coronary arteries. To evaluate the performance of these two attention modules, we remove self-attention from DMQCA model and remove context attention from main-view framework, respectively. The superior experiment results in Table \ref{t1} prove the effectiveness of these two attention mechanisms.

\subsubsection{Comparison with Some Existing Methods.} 
\begin{figure}[htbp]
\centering
\subfigure[]{
\begin{minipage}[t]{0.23\linewidth}
\centering
\includegraphics[width=1in]{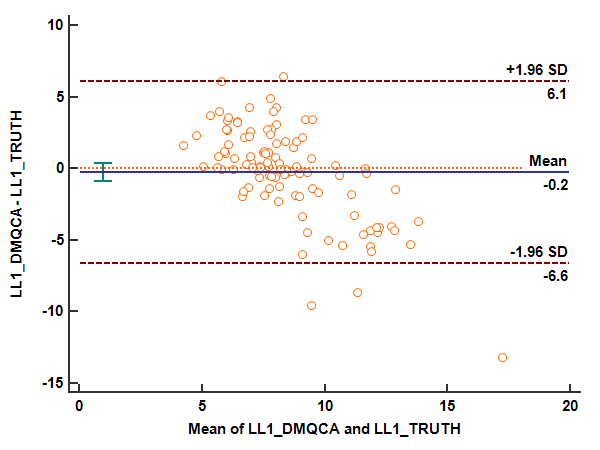}
\end{minipage}%
}%
\subfigure[]{
\begin{minipage}[t]{0.23\linewidth}
\centering
\includegraphics[width=1in]{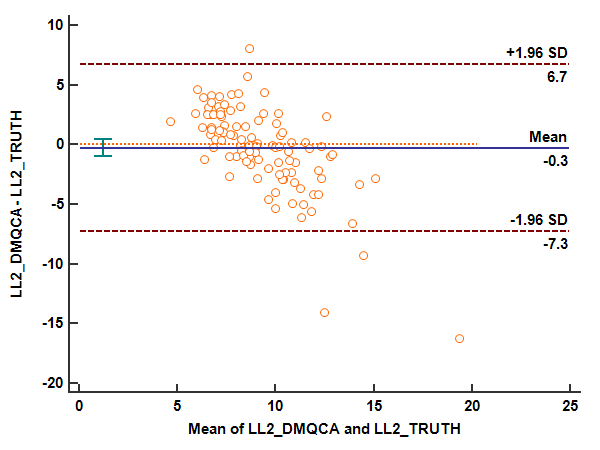}
\end{minipage}%
}%
\subfigure[]{
\begin{minipage}[t]{0.23\linewidth}
\centering
\includegraphics[width=1in]{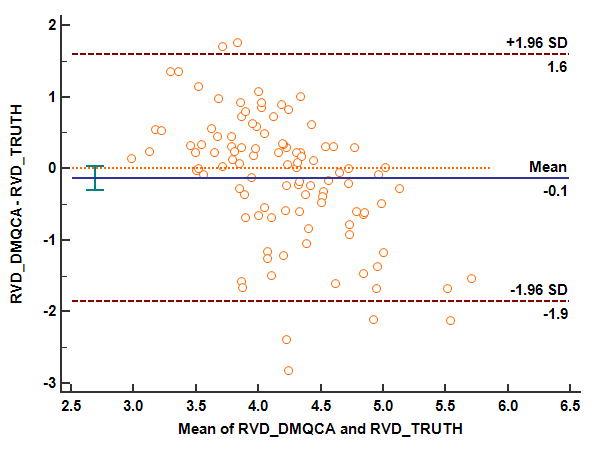}
\end{minipage}%
}%
\subfigure[]{
\begin{minipage}[t]{0.23\linewidth}
\centering
\includegraphics[width=1in]{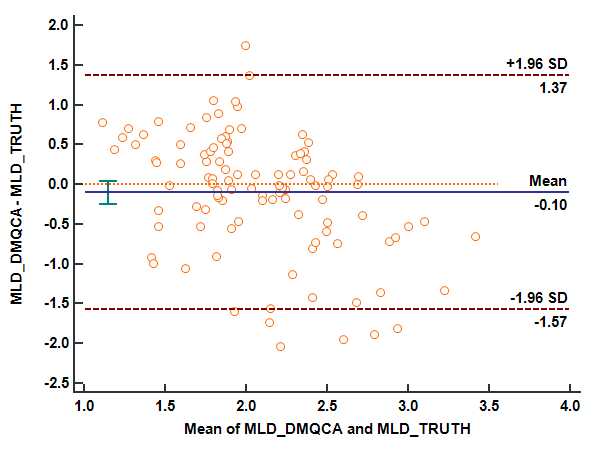}
\end{minipage}%
}%
\centering
\caption{Bland-Altman plots of the quantifications of LL1 (a), LL2 (b), RVD (c), MLD (d).}
\label{fig4}
\end{figure}

In order to evaluate the performance of our DMQCA model, we make the comparison with two baseline methods and three direct quantification methods. As shown in Table \ref{t1}, our proposed DMQCA model has also achieved better quantification performance in the QCA compared with these five methods. We further evaluate the performance of our DMQCA model on different coronary arteries (Fig. \ref{fig3} (a)). And Fig. \ref{fig3} (b) shows example quantification results of different coronary artery lesions obtained by the DMQCA model. In addition, the Bland-Altman plots of indices LL1, LL2, RVD, MLD in Fig. \ref{fig4} (a), (b), (c), (d) indicate the agreement between the DMQCA quantification results and the ground truths.

\section{Conclusion}
In this study, we have proposed a direct quantification model (namely DMQCA) for coronary artery stenosis by incorporating multiview learning and attention mechanisms. By aggregating the multiview features and keyframe feature for direct quantitative index estimation. The proposed method has been evaluated on 105 subjects and has yielded better quantitative results, by comparing with two baseline methods and three existing direct quantification methods. In addition, the experimental results show compelling evidence of our DMQCA method in quantitative coronary angiography.

\subsubsection{Acknowledgments.} This work was supported by the National Natural Science Foundation of China under Grants 61876001, 61602003 and 61673020, the Anhui Provincial Natural Science Foundation under Grant 1708085QF156.


\begin{thebibliography}{10}
\providecommand{\url}[1]{\texttt{#1}}
\providecommand{\urlprefix}{URL }
\providecommand{\doi}[1]{https://doi.org/#1}

\bibitem{zhang2019deep}
Zhang, N., Yang, G., Gao, e.a.: Deep learning for diagnosis of chronic
  myocardial infarction on nonenhanced cardiac cine mri. Radiology
  \textbf{291}(3),  606--617 (2019)

\bibitem{Wan}
Wan, T., Feng, H., Tong, C., Li, D., Qin, Z.: Automated identification and
  grading of coronary artery stenoses with x-ray angiography. Computer Methods
  and Programs in Biomedicine  \textbf{167},  13--22 (2018)

\bibitem{YangJ}
Cong, W., Yang, J., Ai, D., Chen, Y., Liu, Y., Wang, Y.: Quantitative analysis
  of deformable model-based 3-d reconstruction of coronary artery from multiple
  angiograms. IEEE Transactions on Biomedical Engineering  \textbf{62}(8),
  2079--2090 (2015)

\bibitem{chen2018multiview}
Chen, J., Yang, G., Gao, Zhifan, e.a.: Multiview two-task recursive attention
  model for left atrium and atrial scars segmentation. In: MICCAI. pp.
  455--463. Springer (2018)

\bibitem{yang2018multiview}
Yang, G., Chen, J., Gao, Zhifan, e.a.: Multiview sequential learning and
  dilated residual learning for a fully automatic delineation of the left
  atrium and pulmonary veins from late gadolinium-enhanced cardiac mri images.
  In: 40th EMBC. pp. 1123--1127. IEEE (2018)

\bibitem{Zhang}
Zhang~H, Goodfellow~I, M.D.e.a.: Self-attention generative adversarial
  networks. arXiv preprint arXiv: 1805.08318  (2018)

\bibitem{HAN}
Yang, Z., Yang, D., et~al: Hierarchical attention networks for document
  classification. In: Conference of the North American Chapter of the
  Association for Computational Linguistics: Human Language Technologies. pp.
  1480--1489 (2017)

\bibitem{He}
He, K., Zhang, X., Ren, S., Sun, J.: Deep residual learning for image
  recognition. In: IEEE Conference on Computer Vision and Pattern Recognition
  (CVPR) (2016)

\bibitem{Zhen}
Zhen, X., Wang, Z., Islam, A., Bhaduri, M., Chan, I., Li, S.: Direct estimation
  of cardiac bi-ventricular volumes with regression forests. In: MICCAI. pp.
  586--593 (2014)

\bibitem{Xue}
Xue, W., Islam, A., Bhaduri, M., Li, S.: Direct multitype cardiac indices
  estimation via joint representation and regression learning. IEEE
  Transactions on Medical Imaging  \textbf{36}(10),  2057--2067 (2017)

\bibitem{Xue2}
Xue~W, Brahm~G, P.S.e.a.: Full left ventricle quantiﬁcation via deep
  multitask relationships learning. Medical Image Analysis  \textbf{43},
  54--65 (2018)

\end{thebibliography}

\end{document}